\documentclass{aastex62}
\shorttitle{Visibility null as stellar atmosphere probe}
\shortauthors{Armstrong et al.}
\accepted{for publication in JAI, 21 July 2019}

\usepackage[]{amsmath}
\usepackage[]{amssymb}
\usepackage[]{graphicx}

\begin{document}
\title{Interferometric Fringe Visibility Null as a Function of Spatial Frequency:
  a Probe of Stellar Atmospheres}

\correspondingauthor{J.~T.~Armstrong}
\email{tom.armstrong@nrl.navy.mil}

\author{J.~T.\ Armstrong}
\affil{Naval Research Laboratory \\
Remote Sensing Division \\
4555 Overlook Ave.\ SW \\
Washington, DC  20375, USA}

\author{A.~M.\ Jorgensen}
\affil{New Mexico Institute of Mining and Technology \\
801 Leroy Place \\
Socorro, NM 87801, USA}

\author{D.\ Mozurkewich}
\affil{Seabrook Enginering \\
9310 Dubarry Ave. \\
Seabrook, MD 20706, USA}

\author{H.~R. Neilson}
\affil{Department of Astronomy \& Astrophysics \\
University of Toronto \\
50 St.~George St. \\
Toronto, ON, M5S 3H4, Canada}

\author{E.~K.\ Baines}
\affil{Naval Research Laboratory \\
Remote Sensing Division \\
4555 Overlook Ave.\ SW \\
Washington, DC  20375, USA}

\author{H.~R.\ Schmitt}
\affil{Naval Research Laboratory \\
Remote Sensing Division \\
4555 Overlook Ave.\ SW \\
Washington, DC  20375, USA}

\author{G.~T.\ van Belle}
\affil{Lowell Observatory \\
Mars Hill Road \\
Flagstaff, AZ 86001, USA}

\begin{abstract}

  We introduce an observational tool based on visibility nulls in optical spectro-interferometry
  fringe data to probe the structure of stellar atmospheres.  In a preliminary demonstration, we use
  both Navy Precision Optical Interferometer (NPOI) data and stellar atmosphere models to show that
  this tool can be used, for example, to investigate limb darkening.

  Using bootstrapping with either multiple linked baselines or multiple wavelengths in optical and
  infrared spectro-interferometric observations of stars makes it possible to measure the spatial
  frequency $u_0$ at which the real part of the fringe visibility ${\rm Re}(V)$ vanishes.  That
  spatial frequency is determined by $u_0 = B_\perp/\lambda_0$, where $B_\perp$ is the projected
  baseline length, and $\lambda_0$ is the wavelength at which the null is observed.  Since $B_\perp$
  changes with the Earth's rotation, $\lambda_0$ also changes.  If $u_0$ is constant with
  wavelength, $\lambda_0$ varies in direct proportion to $B_\perp$.  Any departure from that
  proportionality indicates that the brightness distribution across the stellar disk varies with
  wavelength via variations in limb darkening, in the angular size of the disk, or both.
  
  In this paper, we introduce the use of variations of $u_0$ with $\lambda$ as a means of probing
  the structure of stellar atmospheres.  Using the equivalent uniform disk diameter
  $\theta_{\rm UD, 0}(\lambda_0)$, given by $\theta_{\rm UD, 0} = 1.22/u_0(\lambda_0)$, as a
  convenient and intuitive parameterization of $u_0(\lambda_0)$, we demonstrate this concept by
  using model atmospheres to calculate the brightness distribution for $\nu$~Ophiuchi and predict
  $\theta_{\rm UD, 0}(\lambda_0)$, and then comparing the predictions to coherently averaged data from
  observations taken with the NPOI.

\end{abstract}


\section{Introduction}\label{sec:intro}

In 1891, Michelson placed a two-slit mask across the aperture of the 12-inch telescope on
Mt.\ Hamilton and used it to observe Jupiter's Galilean satellites.  Light passing through the slits
from the satellite being observed produced interference fringes in the focal plane.  Michelson
widened the slit separation until the fringe contrast (the ``visibility'') went to zero
\citep{Michelson1891}, thereby producing a measure of the satellite's angular diameter: the spatial
frequency at $u_0$ which the fringes vanished, given by $u_0 = \lambda\, /\, S_0$, where $S_0$ is
the slit separation producing zero visibility, and $\lambda$ is the observing wavelength.  Three
decades later, he applied this method to stellar observations at Mt.\ Wilson \citep{Michelson21},
using outrigger mirrors on the 100-inch telescope in place of a mask with slits.

While measuring stellar angular diameters has become one of the primary activities in optical and
infrared interferometry --- results\footnote{For an up to date list, see the Jean-Marie Mariotti
  Center publication database at http://jmmc.fr/bibdb/} over the past quarter century in the
optical, near-IR, and mid-IR include angular measurements of the pulsations of Cepheids
\citep[e.g.,][]{Lane00, Gallenne18} and of their extended envelopes \citep[e.g.,][]{Kervella06}, Be
star disks \citep[e.g.,][]{Mourard89, Gies07}, and debris disks \citep[e.g.,][]{Absil06, Ertel14},
as well as diameter surveys, most recently \citet{Baines18} --- we focus here on developing another
use for $u_0$: using its variation with wavelength as a probe of the stellar atmosphere.

To set the stage for this idea, consider that most interferometric diameter measurements amount to
measuring the fringe visibility amplitude $V$ at one or more spatial frequencies $u$ and using a
model fit, usually a uniform disk,\footnote{Here and below, ``disk'' refers, of course, to the
  appearance of the star rather than to material surrounding it.} to the data to infer the uniform
disk diameter $\theta_{UD}$.  Here, $u = B_{\perp}\, /\, \lambda$, where $B_{\perp}$ is the
component of the interferometer baseline perpendicular to the direction to the star.  Conceptually,
this method is equivalent to extrapolating $V(u)$ to the spatial frequency $u_0$ at which the
visibility of a uniformly bright disk vanishes.  One then calculates the uniform-disk diameter
$\theta_{UD}$ in radians via
\begin{equation}\label{equ:theta_UD}
\theta_{UD} = 1.22\, /\, u_0  .
\end{equation}
Since most stars, later types in particular, exhibit a significant amount of limb darkening,
investigators have usually estimated the actual diameter of the limb-darkened disk, $\theta_{LD}$,
by applying a correction based on models that characterize limb darkening with a few parameters.
In a few cases, a limb-darkened disk model, again generated from a parameterized model, is fit
directly to the data; see \citet{Baines18} for examples.

In principle, $u_0$, and hence $\theta_{UD}$, are functions of $\lambda$; in fact, some studies,
\citep[e.g.,][]{Mozurkewich03a}, have used the differences in $\theta_{UD}$ between spectral bands
as a means of evaluating a limb darkening model.  It is the wavelength dependence of $u_0$ that we
investigate here as a tool for characterizing stellar atmospheres.

\section{The visibility null}\label{sec:vis_null}

Although Michelson found the visibility null itself in his Mt.\ Hamilton and Mt.\ Wilson work, most
interferometric diameter measurements are made at spatial frequencies smaller than $u_0$ and in
effect are extrapolated to estimate $u_0$, as mentioned above.  \citet{Jorgensen10} and
\citet{Armstrong12} called attention to the fact that one can measure, rather than estimate, $u_0$
by observing across the range of $u$ in which $V(u)$, or more precisely, ${\rm Re}[V(u)]$, goes from
positive to negative.\footnote{If the brightness distribution departs from central symmetry,
  ${\rm Re}(V)$ may not have a null.  We consider only {\em circularly} symmetric sources here, for
  which $V$ has a null, and that null does not vary with the position angle.}  We can do so, despite
having no fringe contrast at $u_0$ to enable fringe tracking, if there is enough signal to track
fringes simultaneously at longer wavelengths (where $u < u_0$) (``wavelength bootstrapping''), or if
the fringe on the zero-crossing baseline is stabilized by fringe tracking on shorter baselines
(``baseline bootstrapping'').

\citet{Jorgensen10} showed baseline-bootstrapped examples for three bright stars observed with the
Navy Precision Optical Interferometer \citep[NPOI;][]{Armstrong98a, vanBelle18}.  They fitted
uniform-disk models to the data near the zero crossings and used those fits to interpolate the
wavelength $\lambda_0$ at which ${\rm Re}[V(u)] = 0$, which yields $u_0$ via
\begin{equation}\label{u_0}
u_0 = B_\perp/\lambda_0 .
\end{equation}  

This method recapitulates Michelson: he too sought the null ($V^2 = 0$ in his case, rather than
${\rm Re}[V(u)] = 0$), although he did so by measuring in effect across a range of $B_\perp$ rather
than a range of $\lambda$.  The result in both cases is a measurement of $\theta_{UD}$ via
Eq.~\ref{equ:theta_UD}.  \citet{Jorgensen10} obtained $\theta_{UD}$ results with uncertainties
ranging from 0.8\% to 0.08\%.  This level of precision is possible because finding $u_0$, and hence
$\theta_{UD}$, directly is insensitive to multiplicative errors in visibility calibration.  For
comparison, if one calculates $\theta_{UD}$ using data where $V^2$ is no smaller than $0.4$, a 1\%
error in calibration corresponds to a 0.5\% error in $\theta_{UD}$.

This approach offers the possibility of revealing detailed stellar atmosphere information as a
function of wavelength.  As the Earth rotates, $B_\perp$ also changes.  As a result, $u_0$ also
changes because of its dependence on both $B_\perp$ (Eq.~\ref{u_0}) and any variation in
$\theta_{UD}$ with $\lambda$. It is convenient to parameterize the result as
$\theta_{UD,0}(\lambda_0)$, the uniform disk diameter (``UD'') generated from the visibility zero
crossing (``0'') as a function of the wavelength $\lambda_0$ at which the zero crossing was
observed:
\begin{eqnarray}\label{equ:theta_0_lambda}
\theta_{UD,0}(\lambda_0) & = & 1.22\, /\, u_0(\lambda) \nonumber \\
                        & = & 1.22\, \lambda_0 / B_{\perp} .
\end{eqnarray}
Any variation of $\theta_{UD,0}$ with wavelength is an indication that limb darkening, the effective
diameter of the stellar atmosphere, or both, vary with wavelength as well.  We may not have gained
access to a clear single value of the limb-darkened (LD) diameter, but by treating
$\theta_{UD,0}(\lambda_0)$ as a wavelength-dependent parameter, we have gained a probe of the
stellar atmosphere.

\section{NPOI data} \label{sec:data}

The NPOI \citep{Armstrong98a, vanBelle18} is an optical interferometer located at the Lowell
Observatory site on Anderson Mesa, near Flagstaff, Arizona.  It includes 10 siderostats: four of
them, in fixed locations, form the astrometric subarray, and the other six, which can be moved among
30 stations, form the imaging subarray.  The stations are located along the arms of a Y-shaped feed
system, each arm of which is $\sim 250$~m long.  The array stations provide baseline lengths ranging
from 2.2~m to 437~m, although the baselines that have been commissioned to date range from 9~m to
98~m.  The current array apertures use 12.5~cm of each 50~cm diameter siderostat, but three 1~m
telescopes are currently being installed.

Light is fed from each siderostat or telescope through vacuum feed pipes to the optics laboratory,
where optical path differences (OPDs) between array elements due to array geometry and atmospheric
turbulence are compensated in continuously-variable vacuum delay lines.  In addition to OPD
compensation, the delay lines impose 1~kHz triangle-wave delay dithers on each beam line, which
modulate the OPDs, thereby scanning over the fringe packets.  In the NPOI Classic beam combiner, the
combined beams are dispersed into 16 channels spanning $\lambda\lambda$850--520~nm and detected
synchronously by cooled avalanche photodiodes in Geiger mode.  The VISION beam combiner
\citep{Ghasempour12, Garcia16}, currently being commissioned, will dispense with the delay dither by
using spatial, rather than temporal, modulation.

The usual observing sequence alternates between a scan on a program target and a scan on a nearby
calibrator whose diameter has been measured or can be accurately estimated based on its magnitude
and colors.  Scans are typically 3~min in length.  Depending on the number of stars in the input
observing list and other factors, as many as a dozen scans per night can be taken on a given target.

\begin{figure}[htbp]
\begin{center}
\includegraphics[width=0.8\textwidth,angle=-90,origin=c]{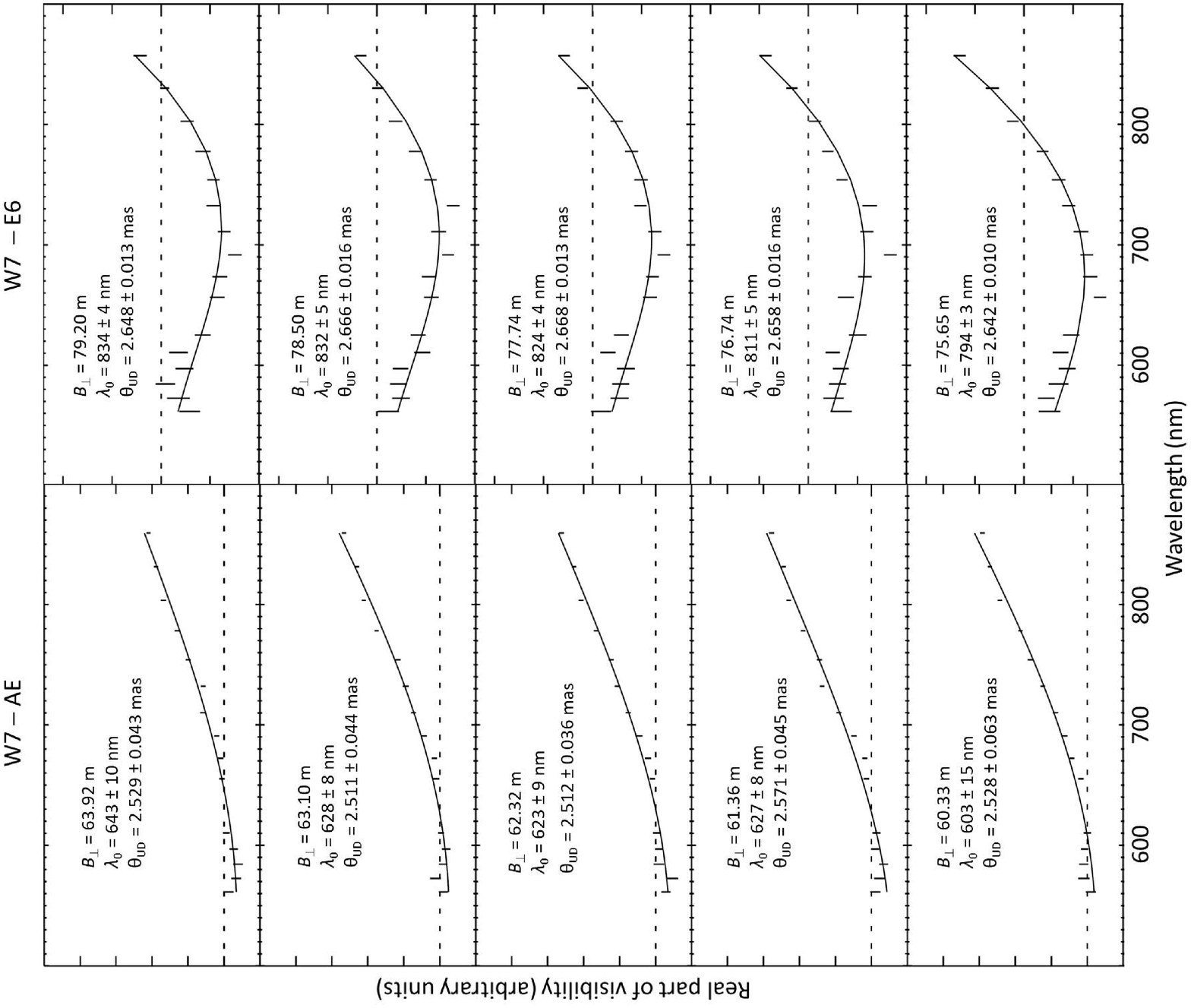}
\end{center}
\caption{Plots of ${\rm Re}[V(\lambda)]$ for five coherently averaged scans of $\nu$~Ophiuchi
  (G9~III) during 2005 June 29 using the bootstrapped W7--AE and W7--E6 baselines of the NPOI.  The
  solid lines represent uniform-disk models multiplied by a wavelength-dependent reduction due to
  phase noise in the data.  Vertical bars show $\pm 1\sigma$ uncertainties.  As the projected
  baseline $B_\perp$ changes with Earth rotation, the wavelength $\lambda_0$ at which ${\rm Re}(V) =
  0$ changes from scan to scan.}
  \label{fig:nu_Oph_ReV}
\end{figure}

We demonstrate our method with five 30~s scans on $\nu$~Ophiuchi (G9~III) taken with the NPOI on
2005 June 29 \citep{Jorgensen10, Armstrong12}.  The observations were taken with the W7, AC, AE, and
E6 stations in 16 spectral channels.  With the AC station as the group delay tracking reference
station, we bootstrapped the 64.4~m W7--AE baseline from the shorter W7--AC and AC--AE
baselines and the 79.4~m W7--E6 baseline from the shorter W7--AC and AC--E6 baselines.
Bootstrapping plus coherent averaging produced high-SNR data at spatial frequencies spanning
${\rm Re}(V) = 0$ on both of these longer baselines.  Figure~\ref{fig:nu_Oph_ReV} shows the
coherently averaged ${\rm Re}[V(\lambda)]$ results.

The curves drawn through the data represent the best-fit uniform disk models multiplied by a
wavelength-dependent reduction factor due to phase noise in the data \citep{Jorgensen10}.  In each
panel, we indicate the projected baseline toward $\nu$~Oph at the time of the observation; in
addition, we show $\lambda_0$ and its uncertainty as estimated from the four to six visibility
measurements straddling $\lambda_0$ and the uniform disk diameter $\theta_{UD,0}$ that produces
${\rm Re}[V(\lambda_0)] = 0$ for baseline length $B_\perp$.  Note that $\sigma(\lambda_0)$ is considerably
greater for data taken on the W7--AE baseline than on the W7--E6 baseline for two reasons: the slope
of ${\rm Re}(V)$ versus $\lambda$ is smaller in the W7--AE data, and the visibilities at the shorter
wavelengths are noisier, as is generally true in NPOI data.

To first order, the difference between $\lambda_0$ values measured on the W7--AE baseline and those
on the W7--E6 baseline are due the difference in $B_\perp$ values: shorter baselines require shorter
wavelengths to resolve a given stellar diameter.  However, the zero-crossing tool, in the form of
the $\theta_{UD,0}$ values calculated from $\lambda_0$ and $B_\perp$, shows that another
effect -- limb darkening in this case -- is also affecting the results.

\section{Model calculations}\label{sec:model_calc}

\begin{figure}[htbp]
\begin{center}
\includegraphics[width=0.8\textwidth]{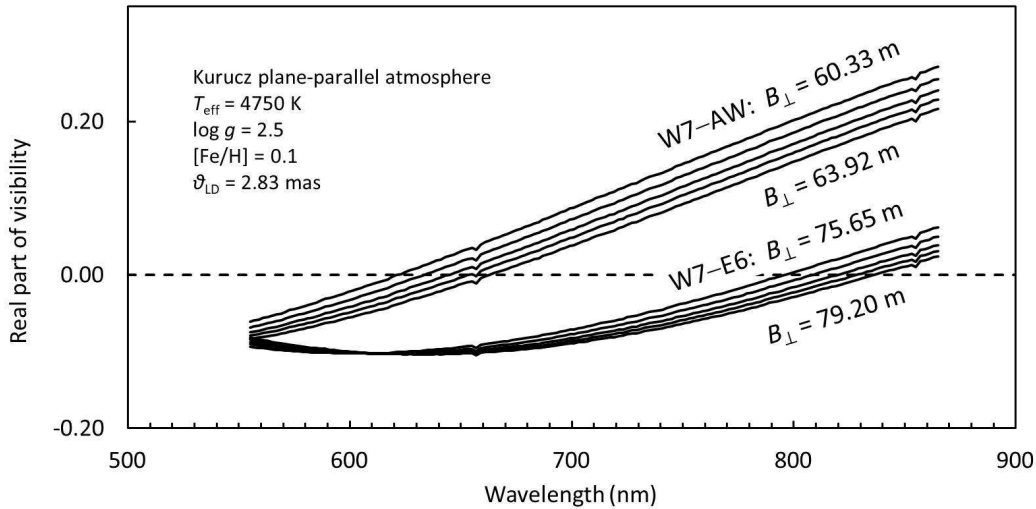}
\end{center}
\caption{Modeled curves of ${\rm Re}[V(\lambda)]$ vs.\ $\lambda$ on two NPOI baselines for
  $\nu$~Ophiuchi based on a Kurucz model plane-parallel atmosphere with $T_{\rm eff} = 4750$~K,
  $\log g = 2.5$, and ${\rm [Fe/H]} = 0.1$.  We calculated ${\rm Re}(V)$ using a diameter
  $\theta_{\rm LD} = 2.83$~mas and with projected baselines $B_\perp$ equal to those used in the
  2005 June 29 observations shown in Fig.~\ref{fig:nu_Oph_ReV}.}\label{fig:nu_Oph_Kurucz_ReV}
\end{figure}

In order to compare these data with the predictions from model atmospheres, we first calculate model
visibilities with the baselines and wavelength range used in the NPOI observations.  Guided by
values for $T_{\rm eff}$ ($4831$~K), $\log g$ (2.7), and [Fe/H] (0.1) from
\citet{Massarotti07,AllendePrieto99} and \citet{McWilliam90}, respectively, we used Kurucz's
brightness profile for a plane-parallel atmosphere {\tt
  ip01k2.pck}\footnote{http://kurucz.harvard.edu/grids.html} (``{\tt p01}" implies
${\rm [Fe/H] = +0.1}$) and extracted the model data for $\log g = 2.50$ and $T_{\rm eff} = 4750$~K.
Kurucz gives the brightness data at 17 values of $\mu$, the cosine of the angle between the normal
to the stellar surface and the line of sight, and for every 2~nm in wavelength in the visual range.
We expanded the grid to 31 values of $\mu$ by linear interpolation.  Selecting an angular diameter
for the star effectively converts the radius $r = \sqrt{1 - \mu^2}$ of an annulus on the stellar
disk to angular units.  For a range of angular diameters, we calculated the Fourier transform at
each wavelength by
\begin{equation}
{\rm Re}[V(\lambda)] = 
    \frac{F(\lambda)}{V_0} 
    \sum_i r_i J_0 \left ( \frac{2\pi B_\perp r_i}{\lambda} \right ) \delta r_i ,
\end{equation}
where $F(\lambda)$ is the intensity at wavelength $\lambda$, $V_0$ is the total flux, which serves
to normalize the visibility to unity at zero baseline, $r_i$ and $\delta r_i$ are the angular radius
and thickness of annulus $i$ on the stellar disk, and $J(\cdot)$ is the Bessel function of the first
kind and zeroth order.  The input diameter that gives the best agreement with the observed
visibility near 800~nm is 2.83~mas (see \S\ref{sec:diameters}).  The resulting model
${\rm Re}[V(\lambda)]$ curves are shown in Fig.~\ref{fig:nu_Oph_Kurucz_ReV}.  Note that the angular
diameter in this procedure is not a uniform disk diameter, but an actual diameter, at least to the
extent that it can reproduce the run of visibilities with wavelength.

\begin{figure}[htbp]
\begin{center}
\includegraphics[width=0.8\textwidth]{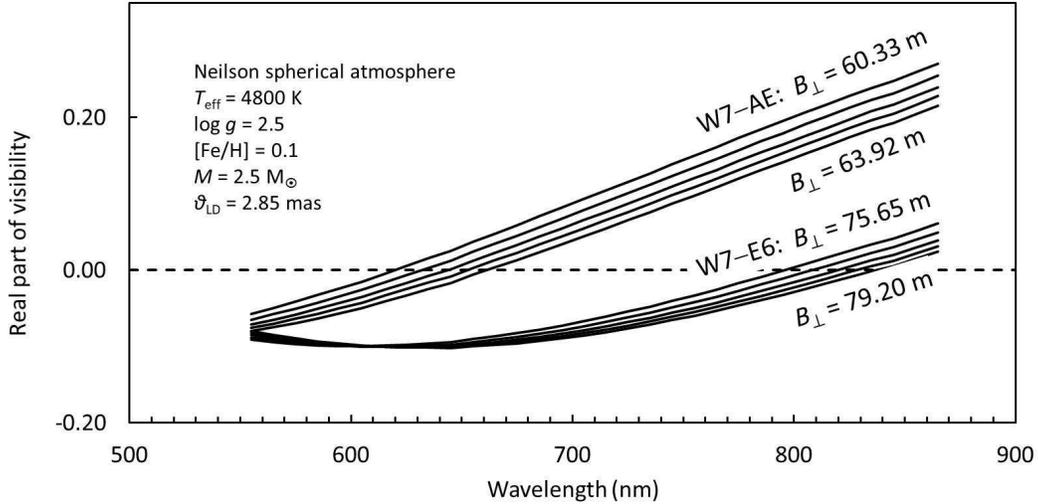}
\end{center}
\caption{Modeled curves of ${\rm Re}[V(\lambda)]$ vs.\ $\lambda$ on two NPOI baselines for
  $\nu$~Ophiuchi based on a Neilson model spherical atmosphere with $T_{\rm eff} = 4800$~K,
  $\log g = 2.5$, $M = 2.5 M_\odot$ and ${\rm [Fe/H]} = 0.1$.  We calculated ${\rm Re}(V)$ using a
  diameter $\theta_{\rm LD} = 2.85$~mas and with projected baselines $B_\perp$ equal to those used
  in the 2005 June 29 observations shown in
  Fig.~\ref{fig:nu_Oph_ReV}.}\label{fig:nu_Oph_Neilson_ReV}
\end{figure}

We applied the same procedure to spherical atmosphere models \citep{Lester08,Neilson08,Neilson11}
calculated by one of us (H.~N.) for $T_{\rm eff} = 4800$~K, $\log g = 2.5$, ${\rm [Fe/H]} = 0.1$,
and mass $M = 2.5 M_\odot$.  The mass, estimated as $2.6 M_\odot$ by \citet{AllendePrieto99}, is a
parameter in these models because it affects the depth of the atmosphere, an effect that is not
important for plane-parallel models.  The Neilson models are given with a wavelength resolution of
10~nm in the visual range and for 1000 values of $\mu$, so no interpolation in $\mu$ was needed.
The resulting model curves are shown in Fig.~\ref{fig:nu_Oph_Neilson_ReV}.  In this case, the input
diameter that best fits the observations near 800~nm is 2.85~mas.

\section{Comparison with NPOI data}\label{sec:diameters}
Figures \ref{fig:nu_Oph_Kurucz_ReV} and \ref{fig:nu_Oph_Neilson_ReV} demonstrate that the
zero-crossing wavelength $\lambda_0$ varies with the projected baseline length $B_\perp$.  Whether
$\theta_{\rm UD,0}$, derived from $\lambda_0$ and $B_\perp$, also varies depends on the brightness
distribution.  If the stellar disk were uniformly bright, $\theta_{\rm UD,0}$ would have a constant
value, regardless of $\lambda_0$ and $B_\perp$.

\begin{figure}[htbp]
\begin{center}
\includegraphics[width=0.8\textwidth]{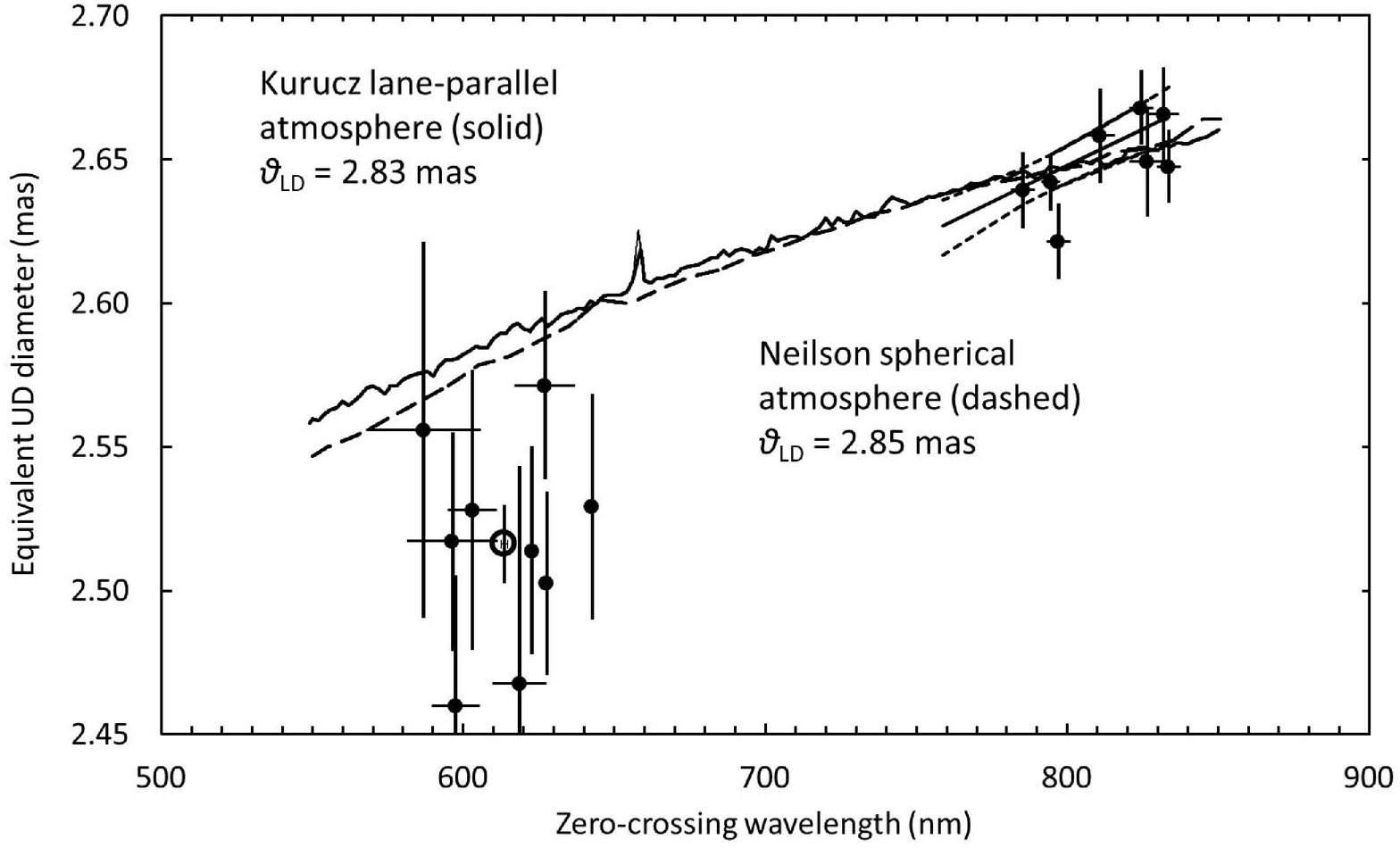}
\end{center}
\caption{Equivalent uniform disk diameters $\theta_{\rm UD,0}(\lambda)$ (filled circles) derived
  from the spatial frequencies $u_0$ of the null in ${\rm Re}(V)$ in NPOI data and compared to
  representative Kurucz and Neilson atmosphere models.  Vertical bars are $1\sigma$ errors.  The
  solid line through the data near $\lambda 800$~nm shows a linear fit to that cluster of points,
  while the short-dashed curves above and below it show the $\theta_{\rm UD,0} \pm \sigma_\theta$
  envelope around the linear fit.  The open circle represents an average of the data in the
  shorter-wavelength cluster.  Also shown are model $\theta_{\rm UD,0}(\lambda)$ curves for the
  Kurucz plane parallel (solid line) and Neilson spherical atmosphere (long-dashed line) models used
  to produce Figs.~\ref{fig:nu_Oph_Kurucz_ReV} and \ref{fig:nu_Oph_Neilson_ReV}, with limb-darkened
  diameters chosen to match the NPOI data near $\lambda 800$~nm.}
  \label{fig:nu_Oph_data+models}
\end{figure}

Not surprisingly, our NPOI data from the G9 giant $\nu$~Ophiuchi demonstrate that the stellar disk
is not uniformly bright.  The dependence of $\theta_{\rm UD,0}$ on $\lambda_0$ is shown in
Fig.~\ref{fig:nu_Oph_data+models}.  The two clusters of $\theta_{\rm UD,0}$ values correspond to
data taken on two baselines.  From a linear fit to the cluster around $\lambda 800$~nm (the solid
line throught those data points in Fig.~\ref{fig:nu_Oph_data+models}), the diameter with the
smallest uncertainty, $\theta_{\rm UD,0} = 2.6475 \pm 0.0042$~mas, occurs at $\lambda_0 = 804$~nm.
The two dashed lines bracketing those data show the envelope of best fit diameters $\pm 1\sigma$.

We have also plotted the predicted runs of $\theta_{\rm UD,0}(\lambda)$ for both a Kurucz and a
Neilson model.  In order to produce agreement with the data near 800~nm, we used limb-darkened
diameters that differ by slightly less than 1\% between the two models: 2.83~mas for the
plane-parallel Kurucz model versus 2.85~mas for the spherical Neilson model.  The difference may be
related to the fact that emission from near the edge of the stellar disk is treated more
realistically by a spherical model.  In addition, the formal ``edge'' of the disk, where $\mu = 1$
at some wavelength, does not represent how the variation of the depth of the atmosphere with
wavelength affects which lines of sight reach an optical depth of unity as they graze the edge of
the disk and which do not.

The curve showing the Kurucz model is actually multi-valued for a small region near the H$\alpha$
line: at projected baseline lengths near 65.13~m, decreased limb darkening near the H$\alpha$ line
due to its higher opacity causes the equivalent uniform-disk diameter to increase.  The Neilson
model does not show this effect because it was calculated on a coarser wavelength grid.  The slopes
of the two models in Fig.~\ref{fig:nu_Oph_data+models} are slightly different; highly precise
zero-crossing measurements would be needed to distinguish between them on this basis.

However, the most important feature of Fig.~\ref{fig:nu_Oph_data+models} is that the observed slope
of $\theta_{\rm UD,0}(\lambda)$ differs significantly from both models, in the sense of requiring
more limb darkening at shorter wavelengths.  Although a detailed study of the atmosphere of
$\nu$~Oph is beyond the scope of this paper, we explored the sensitivity of the slope of the Neilson
model to its input parameters by calculating model $\theta_{\rm UD,0}(\lambda)$ curves for
$T_{\rm eff} = 4800$~K, $\log g = 2.75$, $M = 2.5 M_\odot$ and ${\rm [Fe/H]} = 0.0$ (i.e., slightly
higher $\log g$ and lower metallicity than the model used in Fig.~\ref{fig:nu_Oph_data+models}), and
for $T_{\rm eff} = 4900$~K, $\log g = 2.50$, $M = 2.5 M_\odot$ and ${\rm [Fe/H]} = 0.1$ (slightly
higher $T_{\rm eff}$).  The slopes are virtually identical. Lowering the metallicity and raising the
surface gravity makes a tiny difference. Raising the temperature increases the model diameter by
values ranging from $\approx 0.003$~mas at $\lambda$850~nm to $\approx 0.01$~mas at $\lambda$550~nm.

\citet{Mozurkewich03a} noted the same sense of disagreement between modeled and observed limb
darkening in diameters measured with the Mark~III interferometer, with $\theta_{\rm UD}(800~{\rm
  nm}) / \theta_{\rm UD}(550~{\rm nm}) $ being on average 0.8\% larger than predicted in their
results.  The disagreement in our results for $\nu$~Oph are about the same magnitude.

\section{Conclusion}\label{sec:conclusion}

Inspired by interferometric measurements of stellar angular diameters, we have presented a technique
for measuring a closely related quantity, the uniform-disk diameter calculated from the spatial
frequency at which we observe a null in the fringe visibility.  This technique has the virtue of
being free from multiplicative calibration uncertainties, unlike diameter measurements based on
non-zero visibilities.  As a result, the uncertainties in $\lambda_0$ are due to the number of
photons and the spectral resolution with which they are gathered, and should decline approximately
as $t^{1/2}$.  Each of the scans used in the observations shown here were 30~s long due to
instrumental limitations at the time of the observations, yet the precision with which they
determine $\lambda_0$ is competitive with diameter determinations from nonzero fringe visibilities
with only 1\% calibration errors.  In addition, our data were taken with relatively low spectral
resolution, $\sim 50$.  We expect that the combination of longer integration times and higher
spectral resolution with, e.g., the VISION beam combiner at NPOI, will lead to significantly higher
precision in measuring $\theta_{UD,0}$.

A second virtue of this technique is that we can use the combination of Earth rotation and
spectrally resolved detection to explore how the null moves through spatial frequency space.  In
combination with higher spectral resolution, it will allow us to resolve the effects of such
features as the H$\alpha$ signature apparent in the model curves in
Figs.~\ref{fig:nu_Oph_Kurucz_ReV} and \ref{fig:nu_Oph_data+models}.

The comparison in \S\ref{sec:diameters} shows a distinct disagrement between the data and the
atmosphere models in the sense that the data call for more limb darkening than the models produce.
Our limited exploration of the sensitivity of the models to the input parameters shows that, for
example, changing $T_{\rm eff}$ by 100~K is far from closing the gap.  In future work, we intend to
pursue this discrepancy further and to demonstrate the utility of this observational tool to probe
stellar atmospheres.



\acknowledgments

We thank Robert Kurucz for his assistance in using his atmosphere models and the anonymous review
for comments that significantly clarified this paper.  The Navy Precision Optical Interferometer is
a joint project of the Naval Research Laboratory and the U.S.\ Naval Observatory in partnership with
the Lowell Observatory, and is funded by the Office of Naval Research and the Oceanographer of the
Navy.  This research has made use of the {SIMBAD} database, operated at CDS, Strasbourg, France.


\facilities{Navy Precision Optical Interferometer (NPOI)}


\bibliography{report}
\bibliographystyle{aasjournal}

\end{document}